\documentclass[journal=nanoletters,manuscript=article]{achemso}

\usepackage[version=3]{mhchem} 
\usepackage[T1]{fontenc}
\usepackage[colorlinks=true,citecolor=blue,linkcolor=blue,urlcolor=blue,anchorcolor=blue]{hyperref}%
\hypersetup{colorlinks=true,citecolor=blue,linkcolor=blue,urlcolor=blue}
\providecommand{\U}[1]{\protect\rule{.1in}{.1in}}
\usepackage{amsmath}
\usepackage{amssymb}
\usepackage{graphicx}
\usepackage{mathrsfs}
\usepackage{lineno}
\usepackage{amsfonts}
\usepackage{amsthm}
\usepackage{color}
\usepackage{txfonts}
\usepackage{pifont}

\author{Huanhuan Yang}
\author{Lingling Song}
\author{Yunshan Cao}
\author{Peng Yan}
\email{yan@uestc.edu.cn}
\affiliation[University of
Electronic Science and Technology of China]
{School of Electronic Science and Engineering and State Key Laboratory of Electronic Thin Films and Integrated Devices, University of Electronic Science and Technology of China, Chengdu 610054, China}

\title[An \textsf{achemso} demo]
  {Experimental realization of two-dimensional weak topological insulators}

\abbreviations{IR,NMR,UV}
\keywords{American Chemical Society, \LaTeX}

\begin{document}
\setlength{\baselineskip}{16pt}

\captionsetup[figure]{labelfont={bf},name={Figure},labelsep=period}
\textbf{Abstract:}
 We report the experimental realization of two-dimensional (2D) weak topological insulator (WTI) in spinless Su-Schrieffer-Heeger circuits with parity-time and chiral symmetries. Strong and weak $\mathbb{Z}_2$ topological indexes are adopted to explain the experimental findings that a Dirac semimetal (DSM) phase and four WTI phases emerge in turn when we modulate the centrosymmetric circuit deformations. In DSM phase, it is found that the Dirac cone is highly anisotropic and not pinned to any high-symmetry points but can widely move within the Brillouin zone, which eventually leads to the phase transition between WTIs. In addition, we observe a pair of flat-band domain wall states by designing spatially inhomogeneous node connections. Our work provides the first experimental evidence for 2D WTIs, which significantly advances our understanding on the strong and weak nature of topological insulators, the robustness of flat bands, and the itinerant and anisotropic feature of Dirac cones.

\textbf{KEYWORDS:} \emph{2D weak topological insulator, Dirac cones, flat band}


\section{Introduction}
\maketitle
An enduring interest in condensed matter community is to explore novel phases of matter. The discovery of quantum Hall effect \cite{Klitzing1980} has led to a new phase classification via topological order, which goes beyond Landau's paradigm of phase transition (without spontaneous symmetry breaking). The quantized Hall conductance and its robustness have been successfully explained by ``topology'' (e.g., Chern number), a concept originating from mathematics \cite{Thouless1982}. Since then, the study of topological insulators (TIs) becomes one of the most burgeoning research fields in condensed matter physics \cite{Hasan2010,Qi2011,Chiu2016}. Another milestone in TI research is the realization of quantum spin Hall insulator [two-dimensional (2D) TIs] \cite{Kane2005,Kane20052,Bernevig2006,Konig2007}. Subsequently, the 2D TIs are generalized to three-dimensional (3D) cases \cite{FuTI3D,Moore2007,Roy2009,Fu2007}. Strong and weak 3D TIs both host metallic surface states but with an odd and even number of Dirac points, respectively. Strong 3D TIs have been observed in Bi$_{1-x}$Sb$_x$ and Bi$_2$Se$_3$ \cite{Hsieh2008,Xia2009}. However, the discovery of weak topological insulator (WTI) is more than one decade later \cite{Noguchi2019,Zhang2021}, because the weak topological surface states appear only on particular surfaces and they are sensitive to disorder, which usually are undetectable in real 3D crystals \cite{Yan2012,Rasche2013,Tang2014,Yang2014}. Quantum spin Hall insulator, with helical edge modes spreading over all boundaries, is regarded as a strong TI in 2D systems. It is therefore intriguing to ask if there exist 2D WTIs with robust edge states emerging only along certain boundaries \cite{Jeon2021} and particularly how to realize them in experiments.

Recently, electronic circuits were demonstrated to be an excellent platform to simulate topological band physics \cite{Dong2021}. Utilzing electronic components including resistances, inductors, capacitors, and operational amplifiers, one can study abundant topological phenomena, such as (higher-order) topological insulators \cite{Jia2015,Albert2015,Lee2018,Hofmann2019,Zhu2019,Yyt2021,Yang2021,YWang2020,Imhof2018,Ezawa2018,JBao2019,Yang2020,Song2020,Chen2020}, semimetals \cite{Luo2018,Lu2019,YBYang2019,Islam2020,Lee2020,RLi2020}, and non-Hermitian physics \cite{Ezawa081401,Helbig2020,Liu2020,Zhang2020,Zou2021}, among others \cite{Hadad2018,YWang2019,Kotwal2021,Olekhno2020}. In this Letter, we realize the WTIs in a spinless 2D Su-Schrieffer-Heeger (SSH) circuit under centrosymmetric deformations. By evaluating the strong $\mathbb{Z}_2$ topological index $\nu_0$ and two weak $\mathbb{Z}_2$ topological indexes $\nu_1$ and $\nu_2$, we find a Dirac semimetal (DSM) phase ($\nu_0=1$) and four WTI phases ($\nu_0=0$ and $\nu_1\nu_2=11,10,01,00$). The WTIs obey the following bulk-boundary correspondence: the nontrivial bulk topology indicates the existence of dispersionless edge states, i.e., flat bands. The phase transition between WTIs is mediated by the emerging Dirac-point running in the Brillouin zone (BZ), which is in sharp contrast to conventional cases that the Dirac points are pinned to high-symmetry points of BZ. For circuits with open boundary conditions, we predict edge states exclusively along non-zero $\nu_1\nu_2$ boundaries, which are confirmed by impedance measurements. Flat-band domain wall (DW) states conceived by Zhu \emph{et al.} \cite{ZPA2019} are also designed  and measured by introducing inhomogeneous capacitor connections between nodes.

\section{Results and discussion}

\begin{figure*}
  \centering
  \includegraphics[width=0.96\textwidth]{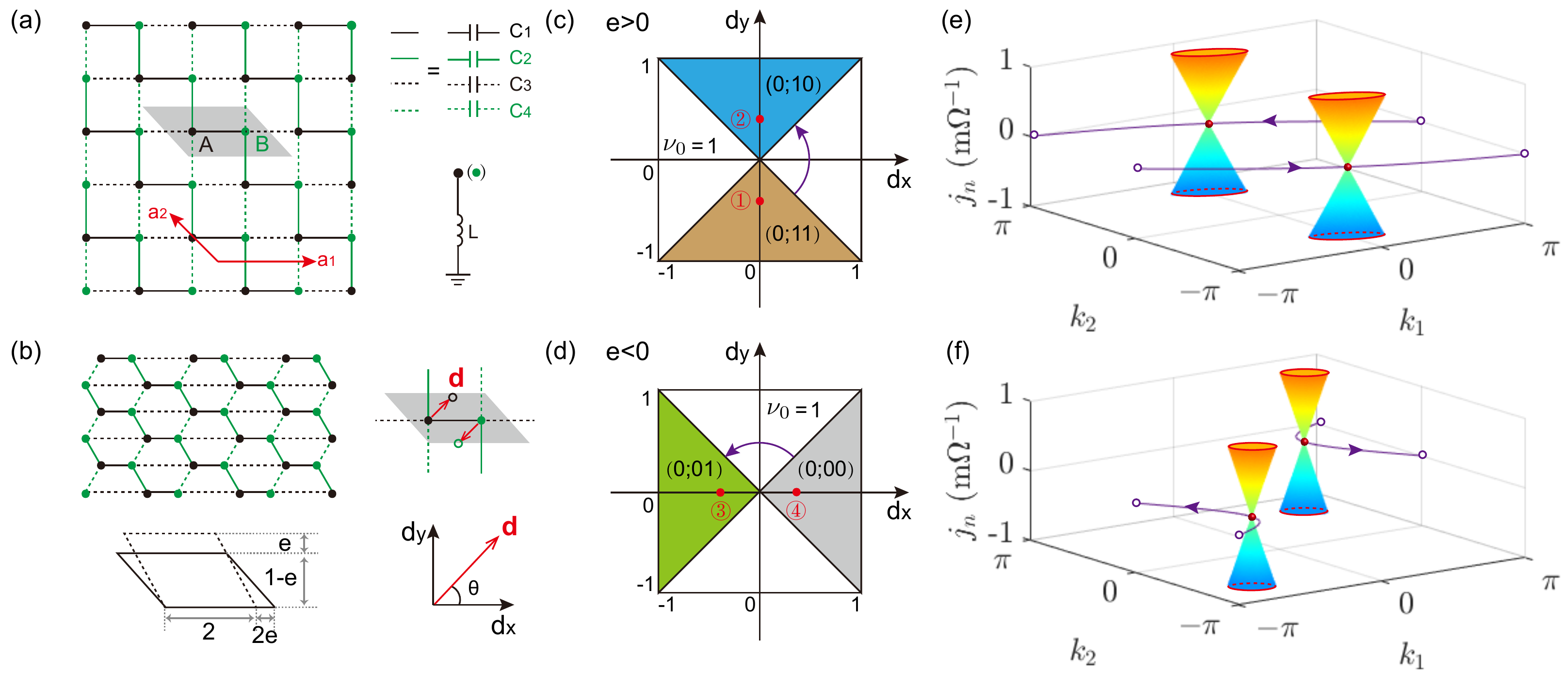}\\
  \caption{(a) Illustration of an undistorted infinite 2D SSH lattice. The segments represent the four kinds of capacitors shown in the inset, and each node is grounded by an inductor $L$. The shaded region displays the unit cell with $A,B$ nodes. ${\bf a}_1=2\hat{x}$ and ${\bf a}_2=-\hat{x}+\hat{y}$ are two basic vectors. (b) Crystals with centrosymmetric deformations $d_x=1/3$, $d_y=0$, and $e=1/3$. Inset: the diagram of  deformation parameters {\bf d} and $e$. (c) and (d) Four WTI phases caused by the crystal deformations, colored by brown, blue, green, and gray, respectively. Topological indexes ($\nu_0$,$\nu_1\nu_2$) are highlighted in each region. (e) and (f) Dirac cone running inside the BZ by continuously tuning the deformation parameters along the purple path in (c) and (d), respectively.}\label{model}
\end{figure*}

We consider an infinite 2D SSH lattice consisting of inductor-capacitor ($LC$) network, as shown in Figure \ref{model}a. The capacitor between two nodes imitates the hopping term in tight-binding model \cite{Jeon2021}. Parameters  $e$ (a uniform square-to-rectangle distortion) and ${\bf d}=(d_x,d_y)$ (a staggered distortion) are used to describe the deformation of the crystal structure (see Sec. I in Supporting Information for details), as depicted in Figure \ref{model}b. The deformation changes the distance between two sites, and thus modulates the hopping strength as $C_1=(1-e+2d_x)C_0$, $C_2=(1+e+2d_y)C_0$, $C_3=(1-e-2d_x)C_0$, and $C_4=(1+e-2d_y)C_0$ with $C_0=1$ nF being the reference capacitor. Labeling the nodes of the circuit by $a=1,2,...$, we can express the circuit response at frequency $\omega$ according to Kirchhoff's law: $I_a(\omega)=\sum_bJ_{ab}(\omega)V_b(\omega)$, with $I_a$ the external current flow into node $a$, $V_b$ the voltage at node $b$, and $J_{ab}$($\omega$) being the circuit Laplacian: $
J_{ab}(\omega)=i\omega\left[-C_{ab}+\delta_{ab}\left(\sum_nC_{an}-\frac{1}{\omega^2L_a}\right)\right].$
Here $C_{ab}$ is the capacitance between nodes $a$ and $b$, $L_a$ is the grounded inductance at node $a$, and the sum is taken over all nearest-neighbor nodes. The circuit Hamiltonian reads
\begin{equation}
\mathcal{H}(\omega)=-iJ(\omega)=\omega\left(
              \begin{array}{cc}
                h_0 & h_1 \\
                h_1^* & h_0 \\
              \end{array}
            \right),
\end{equation}
where $h_0=C_1+C_2+C_3+C_4-1/(\omega^2 L)$ and $h_1=-C_1-C_2e^{i{\bf k}\cdot{{\bf a}_2}}-C_3e^{-i{\bf k}\cdot{{\bf a}_1}}-C_4e^{-i({\bf k}\cdot{{\bf a}_1}+{\bf k}\cdot{{\bf a}_2})}$ with {\bf a}$_1=2(1+e)\hat{x}$ and {\bf a}$_2=-(1+e)\hat{x}+(1-e)\hat{y}$ the two basic vectors, and $\bf k$ being the wave vector. At resonant frequency $\omega_0=1/\sqrt{L(C_1+C_2+C_3+C_4)}$, the diagonal element $h_0$ vanishes. The Hamiltonian $\mathcal{H}(\omega_0)$ is then invariant under the combined parity-time ($\mathcal{PT}$) symmetric operations, i.e.,  $[\mathcal{PT},\mathcal{H}]=0$, where $\mathcal{P}=\sigma_x$ and $\mathcal{T}$ is the complex conjugation with $\mathcal{T}^2=1$ \cite{Dresselhaus2008}. Under the combined $\mathcal{PT}$ symmetry, the Hamiltonian and the periodic part of the Bloch wave function are constrained to be real valued, and as a result both the Berry curvature and Chern number vanish. In such a case, the ${\mathbb Z}_2$-quantized first Stiefel-Whitney number $\nu_c=\frac{1}{\pi}P\oint d{\bf k}\cdot {\bf A}({\bf k})$ can be used to distinguish the topological phases with {\bf A}({\bf k}) the Berry connection and $P$ being the path-ordering operator \cite{AhnCPB2019,Jeon2021}. In the presence of the inversion symmetry: $\mathcal{H}(-{\bf k})=\mathcal{P}\mathcal{H}({\bf k}) \mathcal{P}^{-1}$, one can calculate the ${\mathbb Z}_2$ topological index by evaluating the parity of the occupied band eigenstates at the four time-reversal invariant momenta. Besides, $\mathcal{H}(\omega_0)$ hosts the chiral symmetry due to $\Gamma \mathcal{H} \Gamma^{-1}=-\mathcal{H}$ with $\Gamma=\sigma_z$, so a nontrivial bulk topology will yield the flat edge bands pinned at zero admittance. Here $\sigma_x$ and $\sigma_z$ are Pauli matrices. The admittance spectrum is given by $j_n=\pm\omega_0[C_1^2+C_2^2+C_3^2+C_4^2
+2C_1C_3\cos(k_1)+2(C_1 C_2+C_3 C_4)\cos (k_2)
+2 (C_1 C_4+C_2 C_3) \cos (k_1+k_2)+2C_2 C_4 \cos (k_1+2 k_2)]^{1/2}$ with $k_1={\bf k}\cdot{{\bf a}_1}$ and $k_2={\bf k}\cdot{{\bf a}_2}$.

We first plot the admittance spectrum for different deformation parameters. For $e>0$, we find two gapless DSM phases and two insulating phases, marked by white and colored triangles, respectively, in Figure \ref{model}c. For $e<0$, one can still identify four regions but in different parameter spaces, as shown in Figure \ref{model}d. We adopt the strong ${\mathbb Z}_2$ invariant $\nu_0$ to characterize these phases
\begin{equation}
(-1)^{\nu_0}=\prod_{i=1}^4\delta_i,
\end{equation}
where $\delta_i=\xi(\Gamma_i)$ is the parity eigenvalues at four time-reversal invariant momenta $\Gamma_{i=(n_1n_2)}=\frac{1}{2}(n_1{\bf b}_1+n_2{\bf b}_2)$ ($n_{1,2}=0$ or 1), with ${\bf b}_1$ and ${\bf b}_2$ being the reciprocal-lattice vectors. The $\mathbb{Z}_2$ index $\nu_0$ is computed by $
(-1)^{\nu_0}=\prod_{i=1}^4\delta_i=-{\rm sgn}(d_x-d_y){\rm sgn}(d_x+d_y){\rm sgn}(e)
$ with sgn the sign function (see Sec. II in Supporting Information for details). We obtain $\nu_0=1$ in the all-white regions, which indicates the existence of massless Dirac cones with two-fold degeneracy \cite{Fu2007,AhnCPB2019}, and $\nu_0=0$ for other gapped regions. Furthermore, by introducing the weak topological invariants, i.e., $\nu_1\nu_2$, along {\bf a}$_1$ and {\bf a}$_2$ directions, akin to the treatment of the 3D WTIs
\begin{equation}
(-1)^{\nu_k}=\prod_{n_k=1,n_{j\neq k}=0,1}\delta_{{\mathbf i}=(n_1n_2)},
\end{equation}
with $k=1,2$, we find nonzero weak topological indexes. One can express the topological invariants as $(-1)^{\nu_1}={\rm sgn}(d_x - d_y){\rm sgn}(d_x + d_y)$ and
$(-1)^{\nu_2}=-{\rm sgn}(d_x - d_y){\rm sgn}(e)$. We obtain $(\nu_1\nu_2)=(11)$, $(10)$, $(01)$, and $(00)$ for the four gapped regimes filled by the brown, blue, green, and gray colors, respectively, in Figures \ref{model}c,d (see Sec. II in Supporting Information for details). The phase transition between two WTI phases [e.g., along the purple paths in Figures \ref{model}c,d] occurs by shifting the Dirac cones
\begin{equation}
(k^{\text{DP}}_1,k^{\text{DP}}_2)=
\pm(2{\rm tan}^{-1}\left[\frac{2d_y}{e+1}
\left(\frac{e}{d_x^2-d_y^2}\right)^{\frac{1}{2}}\right],
-2{\rm tan}^{-1}\left[\frac{1}{d_x-d_y}\left(\frac{d_x^2-d_y^2}{e}\right)^{\frac{1}{2}}\right])
\end{equation}
within the BZ, residing on the purple paths in Figures \ref{model}e,f. Open circles at the ends of the trajectories represent the gap opening. Interestingly enough, we find that the flowing Dirac cones are highly anisotropic (see Sec. III in Supporting Information for details). Such phenomena will provide a remarkable opportunity to manipulate the position and ``Fermi'' velocities of Dirac cone, contrasting sharply with conventional observations that Dirac points are isotropic and pinned to high-symmetry points.

\begin{figure}[t!]
  \centering
  \includegraphics[width=0.98\textwidth]{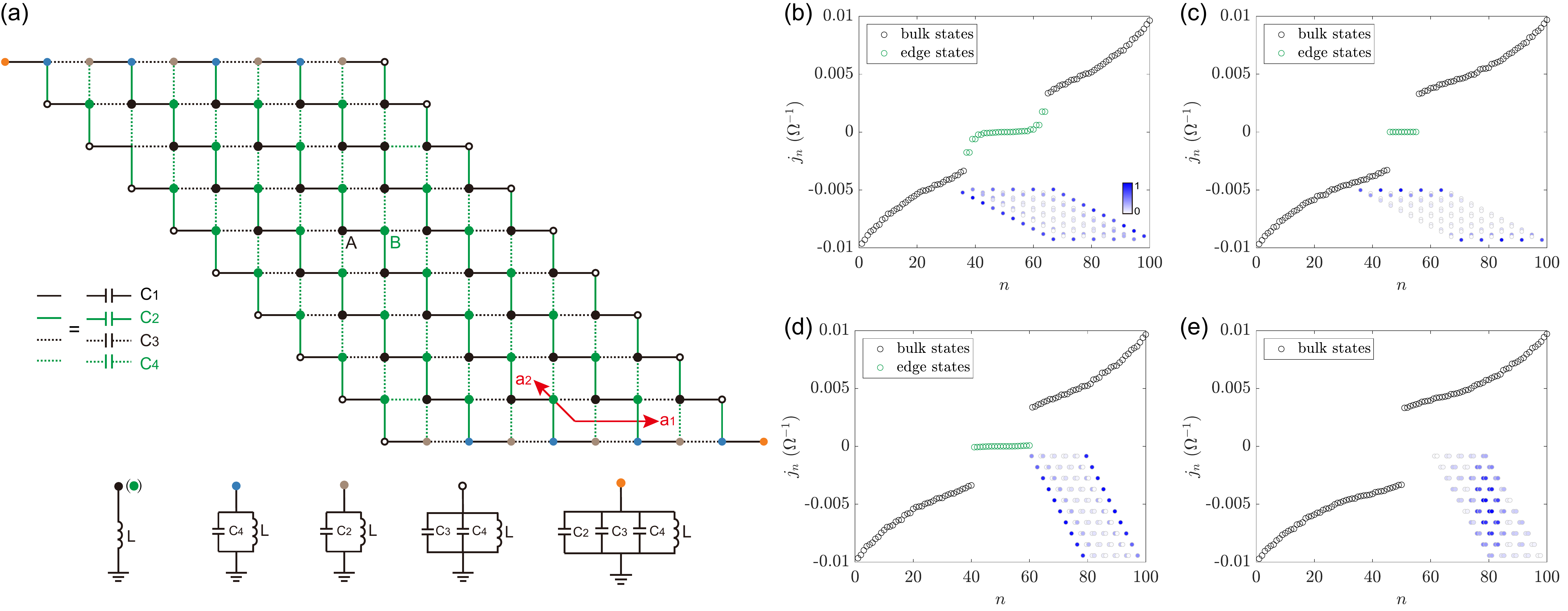}\\
  \caption{(a) A finite-size 2D SSH circuit with 100 nodes. Each node is grounded by inductors and capacitors with the configurations shown in the inset. (b)-(e) Admittance spectrums for four types of parameters: the \ding{172}, \ding{173}, \ding{174}, and \ding{175} red dots shown in Figures \ref{model}c,d. The green and black circles represent the edge and bulk states, respectively.  Insets: the distributions of wave functions with crystal distortions [average of all edge modes for (b)-(d) and the lowest bulk mode for (e)]. }\label{EIG}
\end{figure}

We then consider four finite-size circuits ($\mathcal{N}=100$ nodes) with two edges along ${\bf a}_1$ and ${\bf a}_2$ directions [see Figure \ref{EIG}a]. We adopt $e=0.32$ $(e=-0.32)$, $d_x=0$ $(d_x=\pm0.41)$, $d_y=\pm0.41$ $(d_y=0)$ to realize different deformations [the \ding{172}, \ding{173}, \ding{174}, and \ding{175} red dots in Figures \ref{model}c,d], respectively. Taking $e=0.32$, $d_x=0$, and $d_y=-0.41$ as an example, we choose four kinds of capacitances: $C_1=0.68$ nF, $C_2=2.14$ nF, $C_3=0.68$ nF, and $C_4=0.5$ nF. The other three cases can be realized by exchanging the way of capacitor connections.

By diagonalizing the circuit Laplacians $J_m(\omega)~(m=1,2,3,4)$ for the four cases, we obtain the admittance spectrums and wave functions. Figures \ref{EIG}b-e display the calculation results, where the green and black circles represent the edge and bulk states, respectively. Interestingly, we find edge states existing both in {\bf a}$_1$ and {\bf a}$_2$ directions for parameter \ding{172} [see inset in Figure \ref{EIG}b], which is consistent with the predication of topological index $(\nu_1\nu_2)=(11)$. For the circumstances with $(\nu_1\nu_2)=(10)$ and $(01)$, we observe edge states along ${\bf a}_1$ and ${\bf a}_2$ [see insets in Figures \ref{EIG}c,d], respectively. However, for $(\nu_1\nu_2)=(00)$, only the bulk states can be identified [see inset in Figure \ref{EIG}e]. It's noted that all edge states are located at the zero-admittance flat bands in Figures \ref{EIG}c,d, while only few of them deviate from it [see Figure \ref{EIG}b] due to the hybridization between two edge modes sharing the sample corner.

\begin{figure}[t!]
  \centering
  \includegraphics[width=0.98\textwidth]{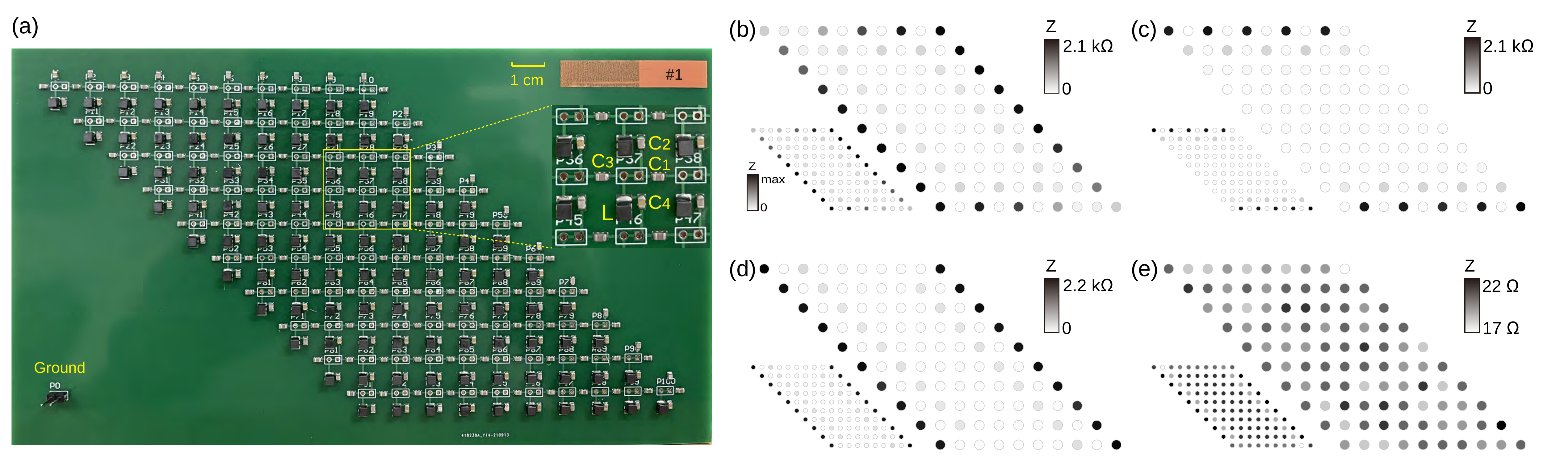}\\
  \caption{ (a) Photograph of one of the four experimental PCBs. The inset zooms in the local details of the circuit. (b)-(e) Measured distributions of the impedance with the same parameters in Figure \ref{EIG}b-e.  Inset: (bottom left) theoretical results.}\label{Impedance}
\end{figure}

To observe the edge states, a convenient method is to measure the spatial distribution of the impedance between each node and the ground \cite{Yang2020,Song2020}. We prepare four printed circuit boards (PCBs) with the same parameters in Figures \ref{EIG}b-e, while each capacitance and inductance allow a 5\% tolerance caused by production process. One photograph of the experimental PCBs is presented in Figure \ref{Impedance}a. We measure the node-ground impedance with the impedance analyzer (Keysight E4990A). We clearly observe the edge states existing for non-vanishing weak topological indexes $(\nu_1\nu_2)=(11)$, $(10)$ and $(01)$, as shown in Figures \ref{Impedance}b-d, but bulk states only for $(\nu_1\nu_2)=(00)$ as plotted in Figure \ref{Impedance}e. Experimental findings are consistent with theoretical predictions [bottom-left insets in Figures \ref{Impedance}b-e].

\begin{figure} [tbp!]
  \centering
  \includegraphics[width=0.7\textwidth]{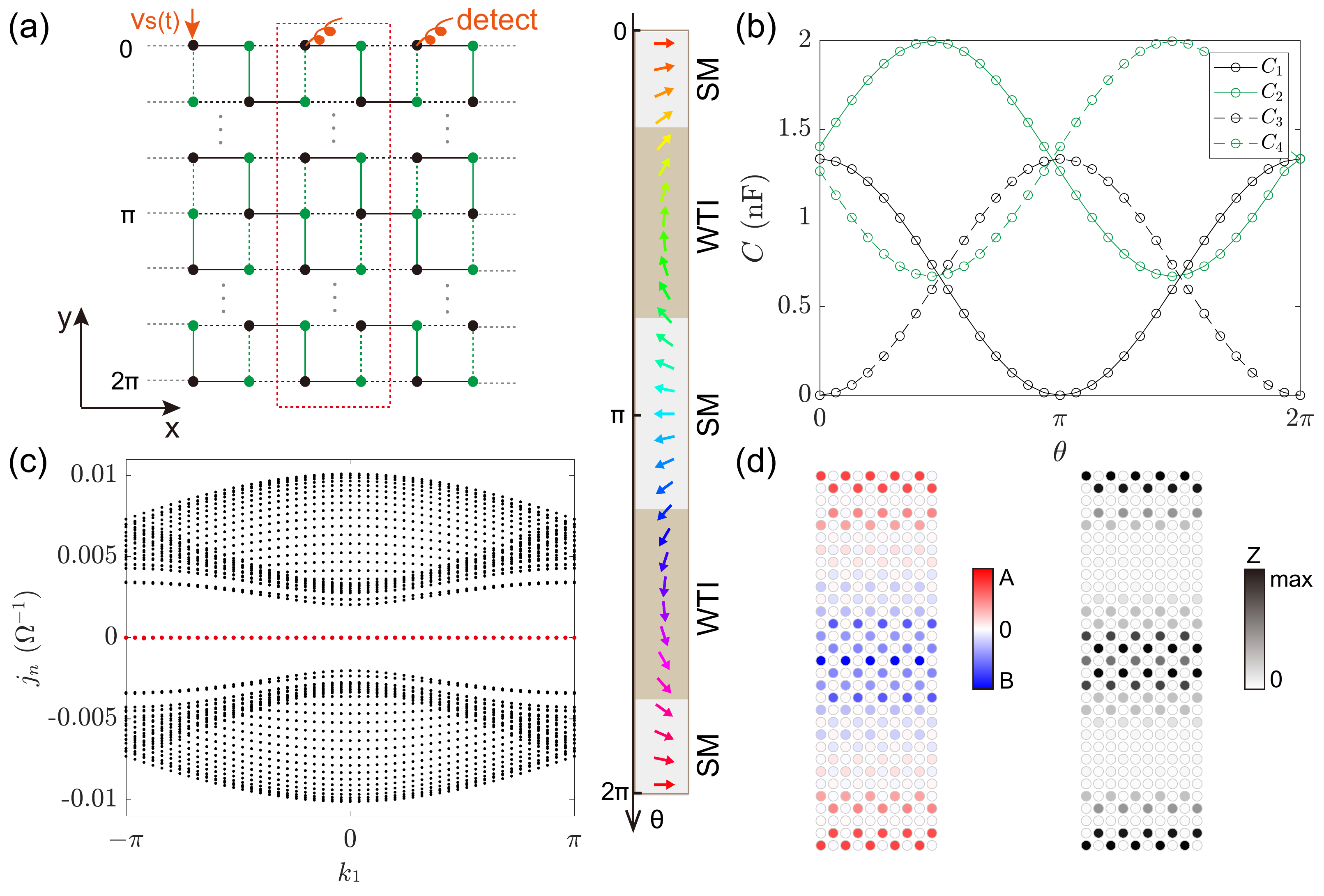}\\
  \caption{(a) Schematic plot of a SSH ribbon with $31$ nodes along $\hat{y}$ direction and periodic boundary conditions in $\hat{x}$ direction. The dashed red rectangle labels the unit cell.  The deformation parameters $e=|{\bf d}|=1/3$, and $\theta$ ranges from 0 to $2\pi$ along $\hat{y}$ direction (see inset). (b) The variation of the hopping ($C_i$) as a function of $\theta$. (c) The numerical admittance spectrum with the zero-admittance flat band (red dots). (d) The distribution of the wave function (DW states) and impedance at $k_1=0$ with 5 unit cells. Red and blue colors represent the intensity of the modes in sublattice $A$ and $B$, respectively.}\label{Ribbon}
\end{figure}

In the above analyses, we focus on the flat-band modes localized at sample edges. Interestingly, if we introduce proper distortions to the lattice, the 1D flat-band bound states can survive in bulk nodes, manifesting as the DW states, which is resembling the zero-dimensional Majorana-like zero modes in distorted Kekul\'{e} lattices \cite{GaoP2019,ChenCW2019} but with one dimension higher. As shown in Figure \ref{Ribbon}a, we construct a DW structure in our circuit by designing the deformation parameter $\theta$ [see Figure \ref{model}b] varying from 0 to $2\pi$ along $\hat{y}$ direction [see inset in Figure \ref{Ribbon}a]. Figure \ref{Ribbon}b displays the four hopping capacitors $C_i$ versus $\theta$ [$C_1=(\frac{2}{3}+\frac{2}{3}\cos\theta)C_0$, $C_2=(\frac{4}{3}+\frac{2}{3}\sin\theta)C_0$, $C_3=(\frac{2}{3}-\frac{2}{3}\cos\theta)C_0$, and $C_4=(\frac{4}{3}-\frac{2}{3}\sin\theta)C_0$]. In principle, one can prepare these very different capacitances by combining several basic capacitors. In this procedure, we fully visit the four phases in Figure \ref{model}c with two DWs in the DSM regions. The topological index $\nu_1$ disappears when $\theta$ goes through the DW regions, which prompts the emergence of zero-admittance flat bands along $k_1$ direction. We plot the admittance in Figure \ref{Ribbon}c, from which one indeed observes a flat band at $j_n=0$. Then, we plot the wave functions of flat band at $k_1=0$ with four-fold degeneracy in the left panel of Figure \ref{Ribbon}d. In the two DSM phases, the mode intensity concentrates on sublattice $A$ and $B$, respectively, manifesting a pair of DW states. In the right panel of Figure \ref{Ribbon}d, we plot the distribution of the impedance, which can well characterize the two DW states, too. In what follows, we present their experimental realizations.

\begin{figure}
  \centering
  \includegraphics[width=0.7\textwidth]{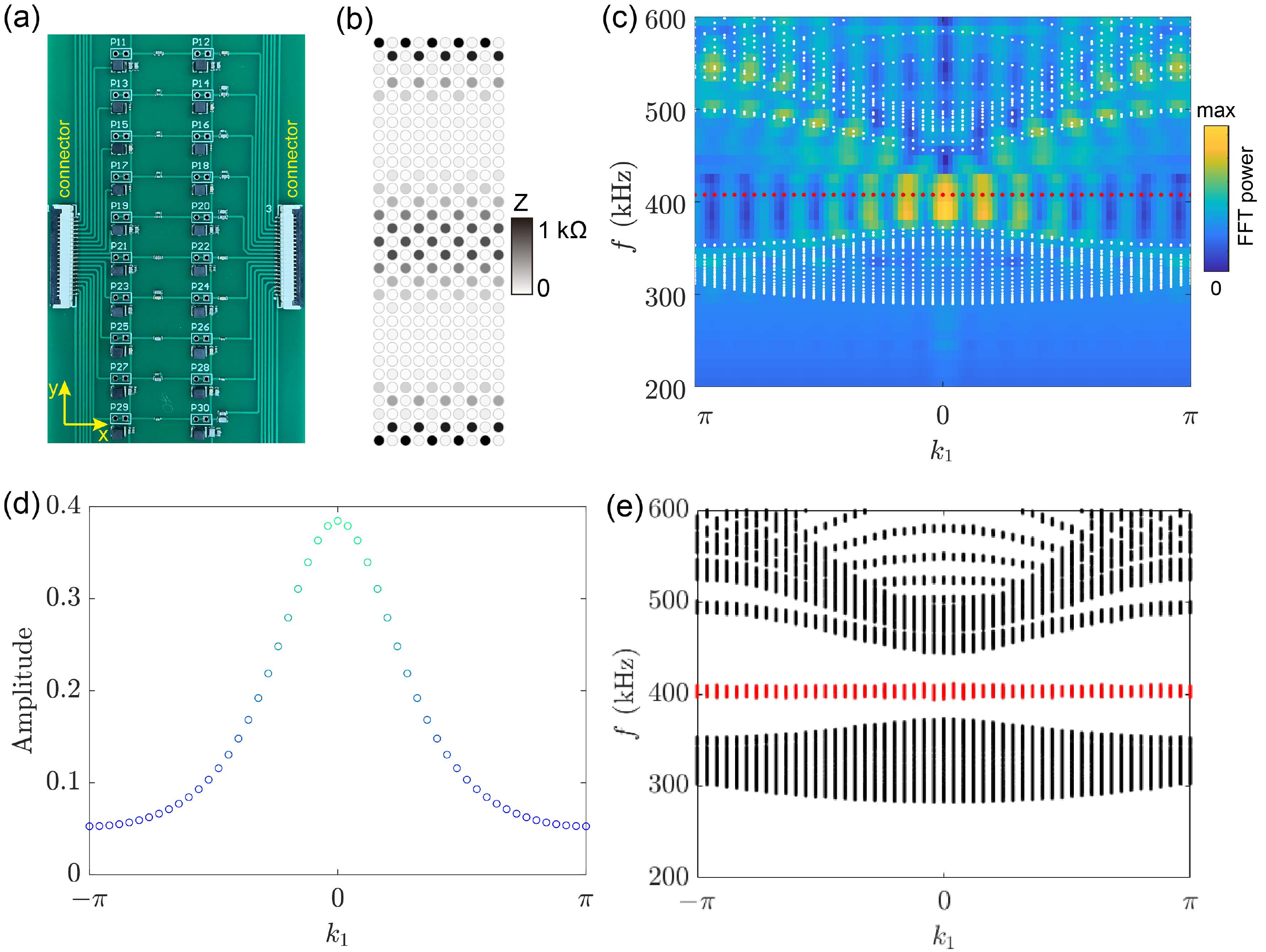}\\
  \caption{(a) The photograph of experimental PCBs (one unit cell). Unit cells are connected to each other by the flat cable connectors. (b) Measured impedance. (c) Frequency spectrum near the resonant frequency. The dots and color map are numerical and experimental results, respectively. (d) The amplitude of the wave functions at the analyzed node dependence on the wave vectors. (e) The numerical result of the frequency spectrum with the elements having 5\% tolerance. }\label{ExpR}
\end{figure}

As shown in Figure \ref{ExpR}a, we manufacture PCBs by combining a sequence of unit cells [the dashed red rectangle in Figure \ref{Ribbon}a]. The measured impedance is plotted in Figure \ref{ExpR}b, which compares well with the numerical results [Figure \ref{Ribbon}d]. In addition, we provide the experimental evidence for the flat band by measuring the frequency spectrum. With the Schr\"{o}dinger-type equation derived from Kirchhoff's law \cite{Yang2021}, we obtain the dispersion spectrum shown by dots in Figure \ref{ExpR}c, where a flat band locates at the resonant frequency $f_0=\omega_0/2\pi=403$ kHz (red dots). To detect the flat band, we produce 15 unit cells along $\hat{x}$ direction in experiment. By inputting a voltage signal $v_s(t)=5 \sin(\omega t)$ V at the boundary node labeled in Figure \ref{Ribbon}a [with the arbitrary function generator (GW AFG-3022)] and measuring the voltages in all cells [with the oscilloscope (Keysight MSOX3024A)], we obtain the voltage distribution $v(\omega,x)$ with frequency $f=\omega/2\pi$ ranging from 200 to 600 kHz. Imposing the Fourier-transformation analysis of $v(\omega,x)$, we obtain the color map in Figure \ref{ExpR}c.  At frequency $f=407$ kHz, we observe a in-gap flat band which clearly demonstrates a vanishing group velocity.

Comparing the numerical and experimental frequency spectrums, we find some inconsistency. Firstly, the measured bulk frequency spectrum is not very clear in the range of $300-350$ kHz. This is because the strength of the spectrum depends on the choice of the analyzed node (see Sec. IV in Supporting Information for details). Secondly, the experimental spectrum is clear at $k_1=0$ but weak away from it. We attributed it to the $k$-dependence of the wave-function amplitude, see Figure \ref{ExpR}d. Finally, we interpret why the flat band suffers a spread in our experimental result. The value of all electric elements used in experiment have a 5\% error, which is the main reason of the large bandwidth of flat band. To illustrate this point, we numerically calculate the frequency spectrum for 100 times with 5\% disorder, with results shown in Figure \ref{ExpR}e, from which one clearly see a broadened flat band as we observed in experiments. Interestingly, one can control the propagation of the DW states by adjusting the band flatness, which can be realized by tuning the hybridization between two DW states (see Sec. V in Supporting Information for details).

\section{Conclusion and Discussion}

To summarize, we reported the experimental observation of WTIs in a spinless 2D SSH circuit respecting $\mathcal{PT}$ and chiral symmetries. The emerging $\mathcal{PT}$ symmetry enables us to adopt the strong and weak $\mathbb{Z}_2$ indexes to characterize the DSM phase and four WTI phases. The phase transition between WTIs was shown to be mediated by shifting the anisotropic Dirac cones within the BZ. We demonstrated that the robust edge states exist only along the non-vanishing weak-indexed edges. The chiral symmetry pins edge modes at zero admittance, forming a flat band. We also realized the flat-band DW states by introducing inhomogeneous circuit capacitor connections.

In condensed matter systems, the Coulomb interaction dominates the flat-band physics such that one can investigate the strong correlation effect, e.g., superconductivity, by delicately twisting bilayer graphene \cite{Bistritzer2011,Cao2018,Balents2020}. Interaction effect in our circuit however can be conveniently explored by engineering the nonlinearities. Engaging the flat bands in circuit platform, we can investigate the Anderson localization with unconventional critical exponents by considering disorder \cite{Goda2006,Chalker2010,LeykamD2013}, and examine many enchanting physical phenomena, such as the wave packet without diffraction \cite{Vicencio2015,Mukherjee2015} and the bosonic condensation \cite{Taie2015,Baboux2016}. By including the (pseduo)spin degree of freedom, we envision the emergence of a weak-type quantum (pseduo)spin Hall effect. One can also explore the properties of the circuit in higher dimensions for realizing other fascinating topological states, such as the nodal-ring and Weyl semimetals \cite{YSu2017,YSu20172,YSu20173}.

In Jeon and Kim's proposal \cite{Jeon2021}, it requires centrosymmetric deformations of the 2D square lattice to realize the 2D WTI, which is extremely challenging from the experimental aspect. In our circuit model, by introducing inductors and capacitors, one can successfully simulate the lattice deformations demanded in the tight-binding modelling. Our strategy thus overcomes a great obstacle for experimentally verifying the very existence of 2D WTI. Through this experimental observation, we establish a complete understanding of the strong and weak insulators in both 2D and 3D. Our findings not only highlight the superiority of the circuit platform but also sets a paradigm to other solid-state systems, including acoustic, photonic, and cold-atom systems, for future studies on fundamental WTIs, itinerant Dirac points, and flat-band DW states.

\section{Associated content}
\subsection{Supporting Information} \label{SI}
The Supporting Information is available free of charge at
https://pubs.acs.org/doi/10.1021/acs.nanolett.xxx.

See Supporting Information for the distortion of the lattices (Sec. I), the parity eigenvalues at time-reversal invariant momenta (Sec. II), the anisotropy of itinerant Dirac cones (Sec. III), the error analysis for the flat-band frequency spectrums (Sec. IV), and the propagation of the domain wall state (Sec. V).

\section{Acknowledgements}
This work was supported by the National Natural Science Foundation of China (Grants No. 12074057, No. 11604041, and No. 11704060).

\section{Abbreviation}
TI, topological insulator; WTI, weak topological insulator; DSM, Dirac semimetal;  2D, two-dimensional; 3D, three-dimensional; DW, domain wall; SSH, Su-Schrieffer-Heeger; BZ, Brillouin zone; LC, inductors and capacitors.

\section{Author information}
\subsection{Corresponding Author}

\textbf{Peng Yan} - School of Electronic Science and Engineering and State Key Laboratory of Electronic Thin Films and Integrated Devices, University of Electronic Science and Technology of China, Chengdu 610054 China;
 Email: yan@uestc.edu.cn

\subsection{Authors}
\begin{flushleft}
\textbf{Huanhuan Yang} - School of Electronic Science and Engineering and State Key Laboratory of Electronic Thin Films and Integrated Devices, University of Electronic Science and Technology of China, Chengdu 610054, China

\textbf{Lingling Song} - School of Electronic Science and Engineering and State Key Laboratory of Electronic Thin Films and Integrated Devices, University of Electronic Science and Technology of China, Chengdu 610054, China

\textbf{Yunshan Cao} - School of Electronic Science and Engineering and State Key Laboratory of Electronic Thin Films and Integrated Devices, University of Electronic Science and Technology of China, Chengdu 610054, China
\end{flushleft}

\subsection{Author Contributions}
Peng Yan conceived the idea and contributed to the project design. Huanhuan Yang and Lingling Song designed the circuits and performed the measurements. Huanhuan Yang developed the theory under the guidance of Peng Yan. Huanhuan Yang and and Peng Yan wrote the manuscript. All authors discussed the results and revised the manuscript.
\subsection{Notes}
The authors declare no competing financial interest.



\end{document}


\setlength{\baselineskip}{16pt}
\renewcommand{\thefigure}{S\arabic{figure}}
\renewcommand{\theequation}{S\arabic{equation}}
\captionsetup[figure]{labelfont={bf},name={Figure},labelsep=period}

\subsection{I. The distortion of circuit lattices}
\begin{figure}[tbp!]
  \centering
  \includegraphics[width=0.95\textwidth]{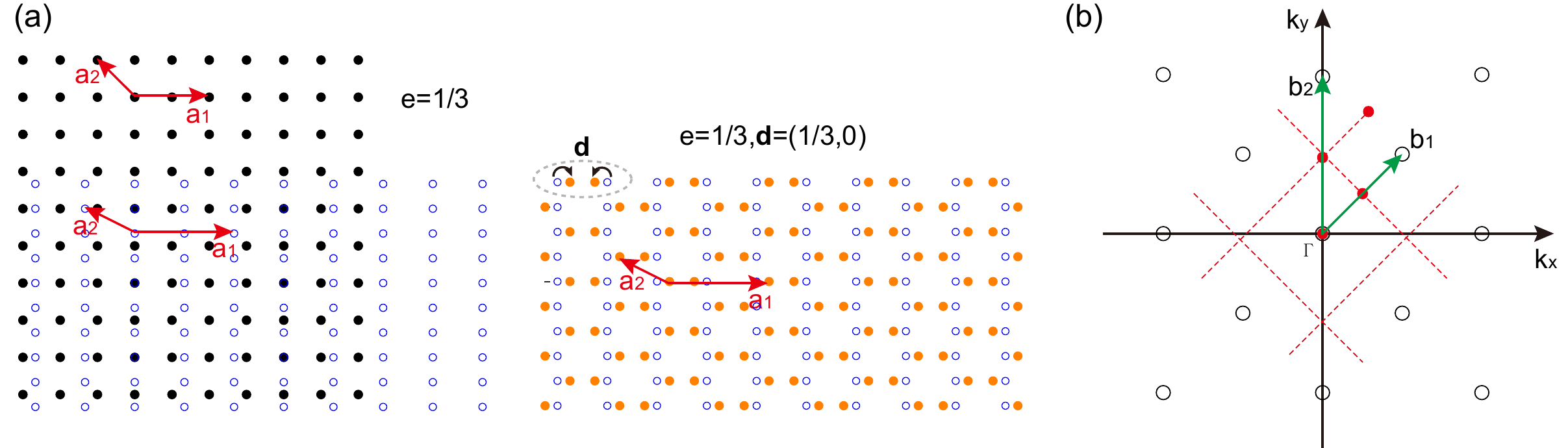}\\
  \caption{ (a) Imposing a uniform square-to-rectangle distortion parametrized by e=1/3 (left panel, black dots to blue circles), and then by staggered distortions parametrized by {\bf d}=($1/3,0$) (right panel, blue circles to orange dots).  (b) The reciprocal lattice with two reciprocal vectors ${\bf b}_1$ and ${\bf b}_2$. The four red dots are TRIM.}\label{BZ}
\end{figure}

In this section, we show the influence of the deformation on the crystal. Before the distortions are introduced [the black dots in Figure \ref{BZ}a], the two basic vectors are chosen as {\bf a}$_1=2\hat{x}$ and {\bf a}$_2=-\hat{x}+\hat{y}$. In Figure \ref{BZ}b, we show their reciprocal vectors ${\bf b}_1=\pi\hat{x}+\pi\hat{y}$, ${\bf b}_2=2\pi\hat{y}$ and four time-reversal invariant momenta (TRIM) $\Gamma_i$ (red dots):
\begin{equation}
\Gamma_{i=(n_1n_2)}=\frac{1}{2}(n_1{\bf b}_1+n_2{\bf b}_2),
\end{equation}
i.e., $\Gamma_1=(0,0); \Gamma_2=(\frac{\pi}{2},\frac{\pi}{2}); \Gamma_3=(0,\pi); \Gamma_4=(\frac{\pi}{2},\frac{3\pi}{2}).$

After we consider the distortions, as shown in Figure \ref{BZ}a, the primitive vectors become {\bf a}$_1=2(1+e)\hat{x}$ and {\bf a}$_2=-(1+e)\hat{x}+(1-e)\hat{y}$. The corresponding reciprocal vectors are given by
\begin{equation}
{\bf b}_1=\frac{\pi}{1+e}\hat{x}+\frac{\pi}{1-e}\hat{y},~
{\bf b}_2=\frac{2\pi}{1-e}\hat{y},
\end{equation}
respectively, with four TRIM:
\begin{equation} \label{Gamma}
\Gamma_1=(0,0); \Gamma_2=\left(\frac{\pi}{2(1+e)},\frac{\pi}{2(1-e)}\right); \Gamma_3=\left(0,\frac{\pi}{(1-e)}\right); \Gamma_4=\left(\frac{\pi}{2(1+e)},\frac{3\pi}{2(1-e)}\right).
\end{equation}

\subsection{II. The parity eigenvalues at time-reversal invariant momenta}

Here, we present the details for the calculation of topological invariants. We use the formula in Ref. \cite{Fu2007}, where the $\mathbb{Z}_2$ is determined by $(-1)^{\nu_0}=\prod_{i=1}^4\delta_i$ where $\delta_i=-{\rm sgn}[d_x({\bf k}=\Gamma_i)]$ with $d_x$ the coefficient of the $\sigma_x$ of the Hamiltonian. The Hamiltonian in the main text is given by:
\begin{equation}
\mathcal{H}(\omega)=\omega\left(
              \begin{array}{cc}
                0 & -C_1-C_2e^{i{\bf k}\cdot{{\bf a}_2}}-C_3e^{-i{\bf k}\cdot{{\bf a}_1}}-C_4e^{-i({\bf k}\cdot{{\bf a}_1}+{\bf k}\cdot{{\bf a}_2})} \\
                -C_1-C_2e^{-i{\bf k}\cdot{{\bf a}_2}}-C_3e^{i{\bf k}\cdot{{\bf a}_1}}-C_4e^{i({\bf k}\cdot{{\bf a}_1}+{\bf k}\cdot{{\bf a}_2})} & 0 \\
              \end{array}
            \right),
\end{equation}
with $C_1=1-e+2d_x$, $C_2=1+e+2d_y$, $C_3=1-e-2d_x$, and $C_4=1+e-2d_y$, which can be rewritten in terms of the Pauli matrixes as:
\begin{equation}\label{H}
\begin{aligned}
\mathcal{H}(\omega)&=-\omega[C_1+C_2\cos({\bf k}\cdot{{\bf a}_2})+C_3\cos({\bf k}\cdot{{\bf a}_1})+C_4\cos({\bf k}\cdot{{\bf a}_1}+{\bf k}\cdot{{\bf a}_2})]\sigma_x\\
&+\omega[C_2\sin({\bf k}\cdot{{\bf a}_2})-C_3\sin({\bf k}\cdot{{\bf a}_1})-C_4\sin({\bf k}\cdot{{\bf a}_1}+{\bf k}\cdot{{\bf a}_2})]\sigma_y.
\end{aligned}
\end{equation}

The parameter $d_x$ is
\begin{equation}
d_x=-\omega[C_1+C_2\cos({\bf k}\cdot a_2)+C_3\cos({\bf k}\cdot a_1)+C_4\cos({\bf k}\cdot a_1+{\bf k}\cdot a_2)].
\end{equation}

At four TRIM ${\bf k}=\Gamma_i$ (Eq. \ref{Gamma}), we obtain:
$d_x^1=-\omega(C_1+C_2+C_3+C_4), d_x^2=-\omega(C_1+C_2-C_3-C_4), d_x^3=-\omega(C_1-C_2+C_3-C_4),$ and $d_x^4=-\omega(C_1-C_2-C_3+C_4).$

\begin{figure}[t!]
  \centering
  \includegraphics[width=0.9\textwidth]{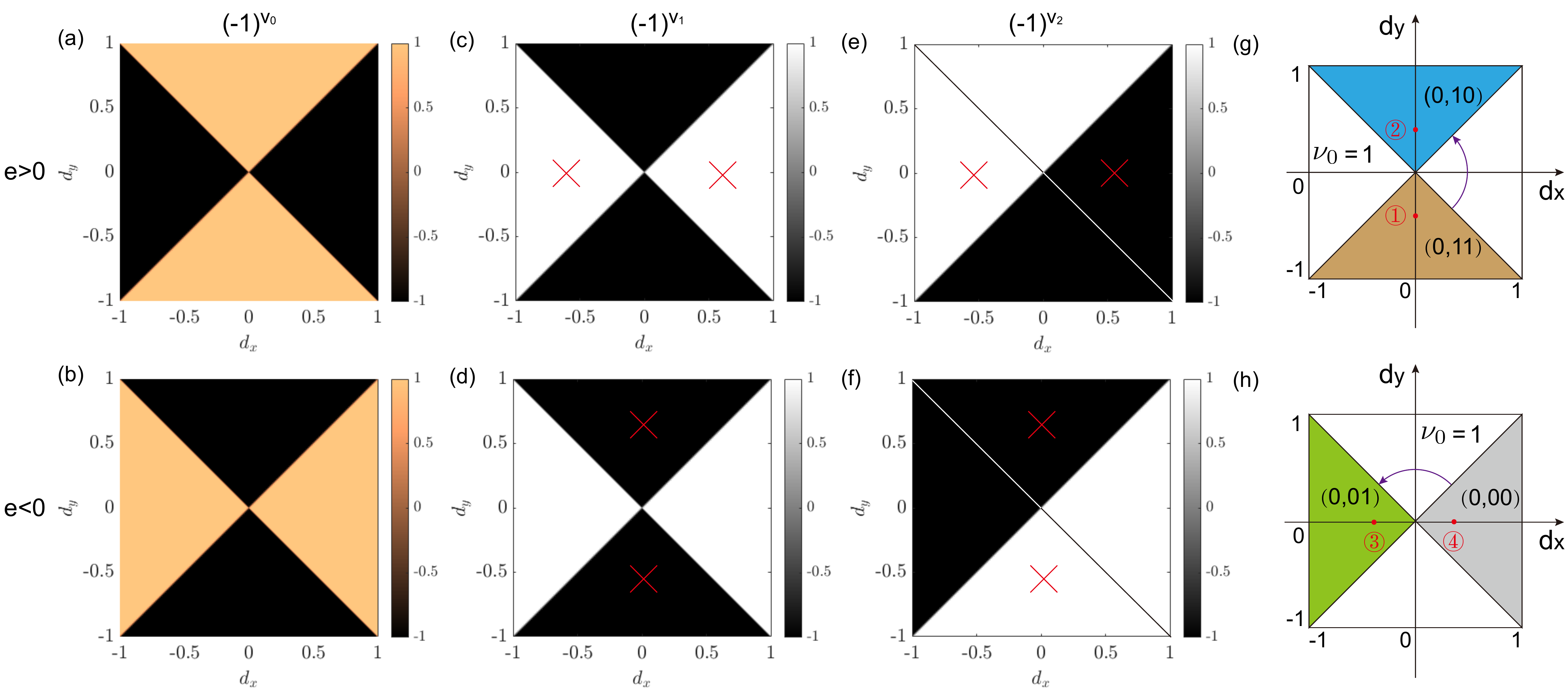}\\
  \caption{The $\mathbb{Z}_2$ indexes $(-1)^{\nu_0}$ for (a) $e>0$ and (b) $e<0$. The two weak topological indexes $(-1)^{\nu_1}$ and $(-1)^{\nu_2}$ for (c)(e) $e>0$ and (d)(f) $e<0$. The red crosses indicate the strong index $\mathbb{Z}_2=1$ (Dirac semimetal regions). Summary of the three topological indexes $(\nu_0,\nu_1\nu_2)$ for (g) $e>0$ and (h) $e<0$.}\label{Z2}
\end{figure}

The $\mathbb{Z}_2$ index $\nu_0$ is
\begin{equation}
(-1)^{\nu_0}=\prod_{i=1}^4\delta_i=-{\rm sgn}(d_x-d_y){\rm sgn}(d_x+d_y){\rm sgn}(e),
\end{equation}
with the parity eigenstates $\delta_i=-{\rm sgn}[d_x^i]$. The results are plotted in Figs. \ref{Z2}a,b.

Next, we focus on the two weak topological indexes:
\begin{equation}
\begin{aligned}
(-1)^{\nu_1}=\prod\delta_{i=(10)}\delta_{i=(11)}={\rm sgn}[d_x^2({\bf k}=\Gamma_2)]{\rm sgn}[d_x^4({\bf k}=\Gamma_4)],\\
(-1)^{\nu_2}=\prod\delta_{i=(01)}\delta_{i=(11)}={\rm sgn}[d_x^3({\bf k}=\Gamma_3)]{\rm sgn}[d_x^4({\bf k}=\Gamma_4)].
\end{aligned}
\end{equation}

Straightforwardly, one can compute $(\nu_1\nu_2)$ as:
\begin{equation}
\begin{aligned}
(-1)^{\nu_1}&={\rm sgn}(d_x - d_y){\rm sgn}(d_x + d_y),\\
(-1)^{\nu_2}&=-{\rm sgn}(d_x - d_y){\rm sgn}(e).
\end{aligned}
\end{equation}
The two weak topological indexes are displayed in Figure \ref{Z2}c-f. Here we do not consider the gapless Dirac simemetal regions, marked by the red crosses.

Finally, we summarize the three topological indexes $(\nu_0,\nu_1\nu_2)$ in Figure \ref{Z2}g for $e>0$ and Figure \ref{Z2}h for $e<0$. The white regions represent the phase $\nu_0=1$, in which a pair of Dirac cones appear. The blue, brown, green, and gray regions indicate the weak TI phases, where the edge states exist in the boundary along ${\bf a}_i$ $(i=1,2)$ directions for nonzero topological indexes.

\subsection{III. The anisotropy of itinerant Dirac cone}

We show that the Dirac cones are highly anisotropic.

\begin{figure}[h!]
  \centering
  \includegraphics[width=0.8\textwidth]{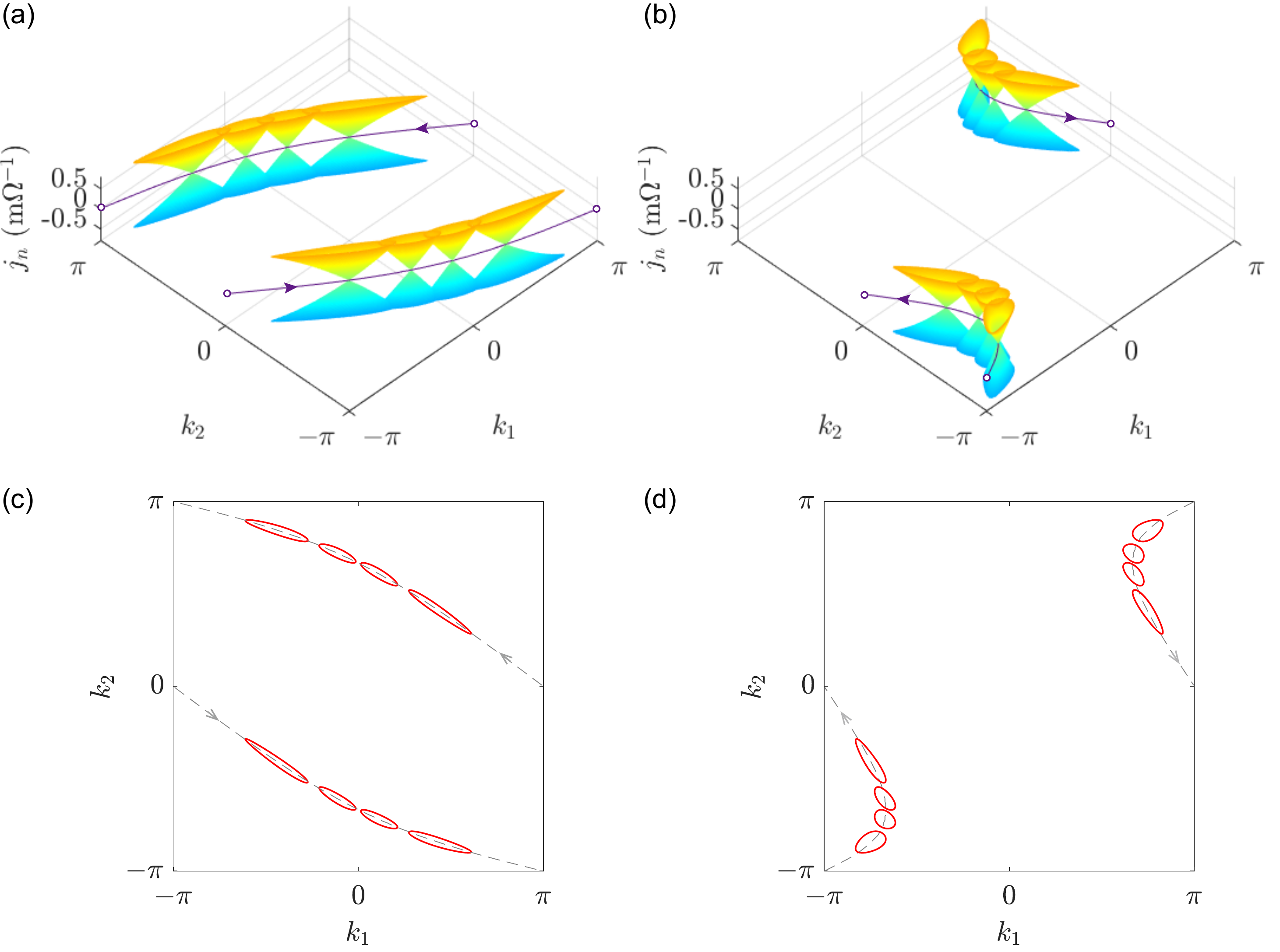}\\
  \caption{ (a)(b) Typical configurations of Dirac cones. (c)(d) The cross-section profiles of the Dirac cones in (a)(b) at $j_n=0.5$ m$\Omega^{-1}$.}\label{DC}
\end{figure}

We begin from the Hamiltonian \eqref{H}:
\begin{equation}
\mathcal{H}(\omega)=-\omega[C_1+C_2\cos(k_2)+C_3\cos(k_1)+C_4\cos(k_1+k_2)]\sigma_x
+\omega[C_2\sin(k_2)-C_3\sin(k_1)-C_4\sin(k_1+k_2)]\sigma_y,
\end{equation}
and obtain the low-energy effective Hamiltonian near the Dirac points as:
\begin{equation}
\begin{aligned}
\mathcal{H}_{\rm eff}=\mathcal{H}(k^{\rm DP}_1,k^{\rm DP}_2)+\frac{\partial \mathcal{H}}{\partial k_1}\bigg|_{(k^{\rm DP}_1,k^{\rm DP}_2)}q_1+\frac{\partial \mathcal{H}}{\partial k_2}\bigg|_{(k_1^{\rm DP},k_2^{\rm DP})}q_2
=(q_1,q_2)\left(
             \begin{array}{cc}
               c_{11} & c_{12} \\
               c_{21} & c_{22} \\
             \end{array}
           \right)\cdot(\sigma_x,\sigma_y)
\end{aligned}
\end{equation}
where $q_1=k_1-k^{\rm DP}_1$ and $q_2=k_2-k^{\rm DP}_2$ are wave vectors near the Dirac points, and $c_{11}=\omega[C_3\sin(k^{\rm DP}_1)+C_4\sin(k^{\rm DP}_1+k^{\rm DP}_2)],
c_{12}=\omega[-C_3\cos(k^{\rm DP}_1)-C_4\cos(k^{\rm DP}_1+k^{\rm DP}_2)], c_{21}=\omega[C_2\sin(k^{\rm DP}_2)+C_4\sin(k^{\rm DP}_1+k^{\rm DP}_2)],
c_{22}=\omega[C_2\cos(k^{\rm DP}_2)-C_4\cos(k^{\rm DP}_1+k^{\rm DP}_2)]$ with $C_1=1-e+2d_x$, $C_2=1+e+2d_y$, $C_3=1-e-2d_x$, and $C_4=1+e-2d_y$.  It's noted that the coefficients of $q_1$ and $q_2$ are different and decided by the distortion parameters {\bf d} and $e$. As a result, the shape of Dirac cones is highly anisotropic and their position is not fixed but flowing within the BZ. Typical configurations of Dirac cones with their cross-section profiles are plotted in Figure \ref{DC} (parameters are adopted from purple paths of Figure 1(e) and 1(f) in the main text), and one clearly see anisotropic Dirac cones.

%

\subsection{IV. The error analysis for the flat-band frequency spectrums}

\begin{figure}[h!]
  \centering
  \includegraphics[width=0.7\textwidth]{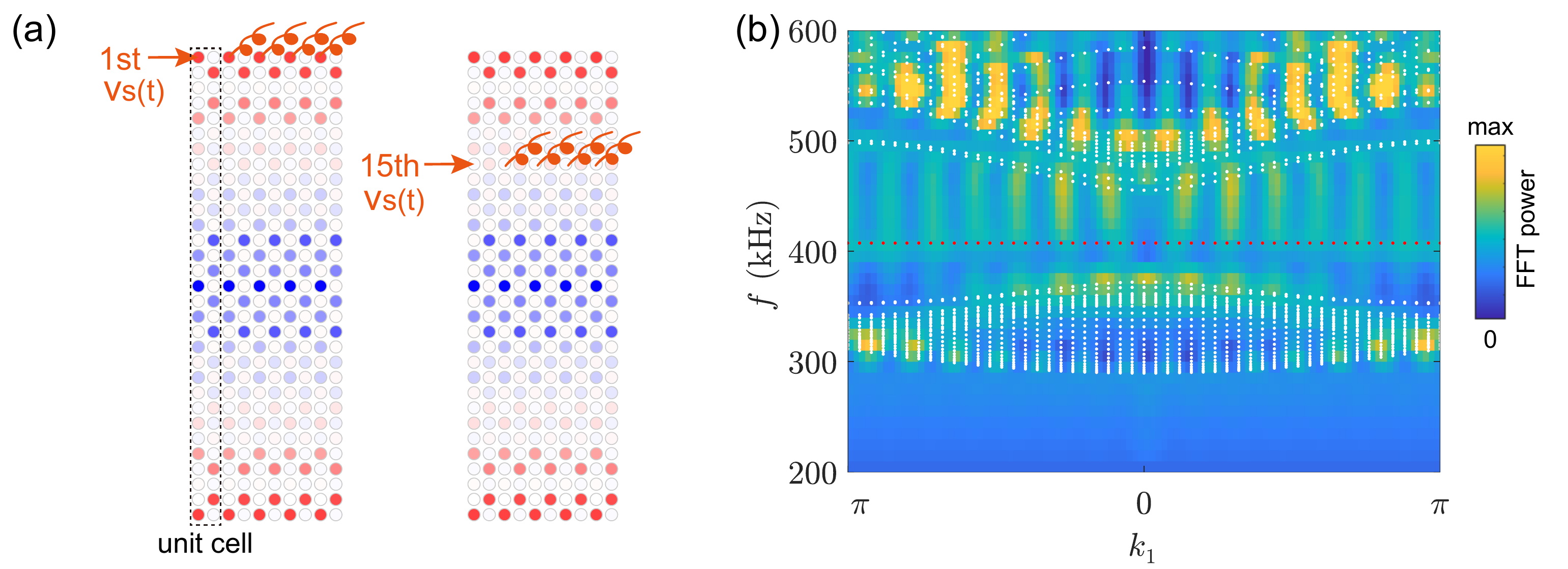}\\
  \caption{ (a) Schematic plot of a ribbon with the signal source and detected positions localized at the 1st and 15th node, respectively. (b) The experimentally measured frequency spectrum. }\label{EFB}
\end{figure}

In this section, we analyze the errors between the numerical and experimental results of the flat band. Firstly, we find the intensity of experimental frequency spectrum in the range $300-350$ kHz is weak. This is because the strength of the spectrum depends on the choice of the analyzed node.  To obtain the spectrum of the flat band, we chose the 1st node of each unit cell of the sample (left panel in Figure \ref{EFB}a), because the 1st node harbors a large wave-function amplitude for the flat band. To display the spectrum in the range of 300 kHz-350 kHz, we change the analyzed point to the 15th one (right panel in Figure \ref{EFB}a). We performed additional measurements with results plotted in Figure \ref{EFB}b, from which one can identify the bulk spectrum in the whole frequency range. However, the flat band is absent, because the wave functions of the flat band hardly locate at the 15th node.

\subsection{V. The propagation of the domain wall state}

\begin{figure}[htbp!]
  \centering
  \includegraphics[width=0.7\textwidth]{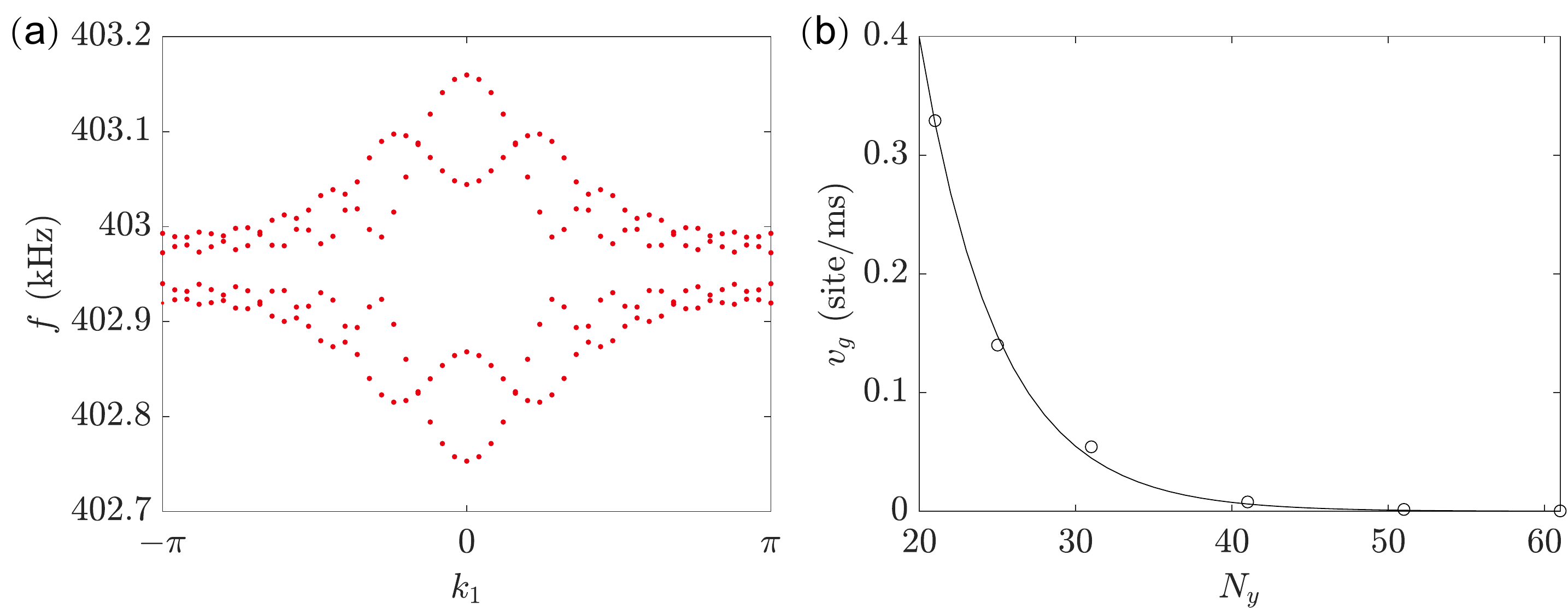}\\
  \caption{(a) The details of the frequency spectrum near the resonant frequency. (b) The estimated group velocity of the DW states. $N_y$ is the number of the site in $\hat{y}$ direction, and the distance between two DWs is $N_y$/2.}\label{vg}
\end{figure}

Here we demonstrate that the propagation of domain wall (DW) states can be controlled by adjusting the flatness of the band. As shown in Figure \ref{vg}a, we enlarge the flat bands of Figure 5c in the main text, and find a small group velocity for the localized state, the reason of which is the two DW states hybridization with each other (like the ones in 1D Su-Schrieffer-Heeger model \cite{Asboth2015}). One can control the group velocity by changing the distance between the DWs. We estimate the group velocity as $v_g=\{\max[{f(k_1=0)}]-\max[{f(k_1=-\pi)}]\}/\pi$ among the flat bands, as displayed in Figure \ref{vg}b. The group velocity of the DW states increases quickly as the distance between two DWs decreases.

